\documentclass{article}
\usepackage{spconf,amsmath,amsfonts,graphicx,hyperref,cite}
\usepackage{array,booktabs,tabularx,colortbl,xcolor}
\hypersetup{colorlinks=true}
\newcolumntype{C}{>{\centering\arraybackslash}X}
\makeatletter
\newcommand\footnoteref[1]{\protected@xdef\@thefnmark{\ref{#1}}\@footnotemark}
\makeatother
\newcommand{\first}[1]{\multicolumn{1}{>{\columncolor[rgb]{1.0,0.7,0.7}}c}{#1}}
\newcommand{\second}[1]{\multicolumn{1}{>{\columncolor[rgb]{1.0,0.85,0.7}}c}{#1}}
\newcommand{\third}[1]{\multicolumn{1}{>{\columncolor[rgb]{1.0,1.0,0.7}}c}{#1}}
\newcommand{\firstl}[1]{\multicolumn{1}{>{\columncolor[rgb]{1.0,0.7,0.7}}l}{#1}}
\newcommand{\secondl}[1]{\multicolumn{1}{>{\columncolor[rgb]{1.0,0.85,0.7}}l}{#1}}
\newcommand{\thirdl}[1]{\multicolumn{1}{>{\columncolor[rgb]{1.0,1.0,0.7}}l}{#1}}
\newcommand{\textstrong}[1]{\textcolor{magenta}{#1}}
\title{MeanVoiceFlow:\\
  One-step Nonparallel Voice Conversion with Mean Flows}
\name{Takuhiro Kaneko, Hirokazu Kameoka, Kou Tanaka, Yuto Kondo}
\address{NTT, Inc., Japan}

\begin{document}
\ninept

\maketitle

\begin{abstract}
  In voice conversion (VC) applications, diffusion and flow-matching models have exhibited exceptional speech quality and speaker similarity performances. However, they are limited by slow conversion owing to their iterative inference. Consequently, we propose \textit{MeanVoiceFlow}, a novel \textit{one-step} nonparallel VC model based on mean flows, which can be trained \textit{from scratch} without requiring pretraining or distillation. Unlike conventional flow matching that uses instantaneous velocity, mean flows employ average velocity to more accurately compute the time integral along the inference path in a single step. However, training the average velocity requires its derivative to compute the target velocity, which can cause instability. Therefore, we introduce a \textit{structural margin reconstruction loss} as a \textit{zero-input constraint}, which moderately regularizes the input--output behavior of the model without harmful statistical averaging. Furthermore, we propose \textit{conditional diffused-input training} in which a mixture of noise and source data is used as input to the model during both training and inference. This enables the model to effectively leverage source information while maintaining consistency between training and inference. Experimental results validate the effectiveness of these techniques and demonstrate that \textit{MeanVoiceFlow} achieves performance comparable to that of previous multi-step and distillation-based models, even when trained \textit{from scratch}.\footnote{\label{foot:samples}Audio samples are available at \url{https://www.kecl.ntt.co.jp/people/kaneko.takuhiro/projects/meanvoiceflow/}.}
\end{abstract}


\section{Introduction}
\label{sec:introduction}

Voice conversion (VC) converts one voice to another while preserving the underlying linguistic information, and has been widely studied to enrich speech communication.
Existing studies have largely focused on parallel VC, as it is simple to train.
However, recent research efforts have shifted toward nonparallel VC because of its greater practical applicability and the fewer constraints on its training data.
In addition, recent advances in deep generative models (e.g.,~\cite{DKingmaICLR2014,DRezendeICML2014,IGoodfellowNIPS2014,OAaronNIPS2017,LDinhICLRW2015}) have led to significant breakthroughs (e.g., \cite{CHsuIS2017,HKameokaTASLP2019,TKanekoEUSIPCO2018,HKameokaSLT2018,KQianICML2019,YHChenICASSP2021,JSerraNeurIPS2019}) that address the challenges inherent to training nonparallel VC as a consequence of the absence of supervised paired data.

In particular, diffusion models~\cite{JSohlICML2015,YSongNeurIPS2019,JHoNeurIPS2020} have recently garnered attention across various research fields, and their applications to VC~\cite{HKameokaTASLP2024,SLiuASRU2021,VPopovICLR2022,HYChoiIS2023,HYChoiAAAI2024} have demonstrated strong speech quality and speaker similarity.
However, a notable limitation of diffusion-based VC is its slow conversion speed, resulting from iterative denoising steps.
Consequently, flow matching models~\cite{YLipmanICLR2023,XLiuICLR2023,MAlbergoICLR2023} and their applications to VC~\cite{JYaoAAAI2025,HYChoiICASSP2025,JZuoICASSP2025,PRenIS2025,HKameokaTASLP2025} have been proposed.
These models learn vector fields that define flow paths between two probabilistic distributions.
This formulation yields straighter trajectories than those in diffusion models, thereby improving sampling efficiency.
However, the performance still degrades significantly when the number of sampling steps is reduced, particularly in the \textit{one-step} case.

Knowledge distillation is a promising approach to this problem in which a multi-step teacher model (e.g., diffusion or flow matching model) is trained and subsequently distilled into a one-step student model (e.g.,~~\cite{TSalimansICLR2022,YSongICML2023,YSongICLR2024}).
Some studies~\cite{ASauerECCV2024,ASauerSIGGRAPHAsia2024} combine knowledge distillation with adversarial training~\cite{IGoodfellowNIPS2014} to enhance performance.
In VC, similar methods have also been explored~\cite{TKanekoIS2024,TKanekoIS2025,TKanekoIS2025b}, and have demonstrated that the performance exhibited by this combination can match that of multi-step models.
However, this approach has two drawbacks:
(1) \textit{increased training cost}, as it requires training both teacher and student models, and
(2) \textit{training instability}, as adversarial training is prone to divergence and often requires a pretrained feature extractor (e.g., a pretrained HiFi-GAN vocoder~\cite{JKongNeurIPS2020}) for stabilization.

\begin{figure}[t]
  \centering
  \centerline{\includegraphics[width=\columnwidth]{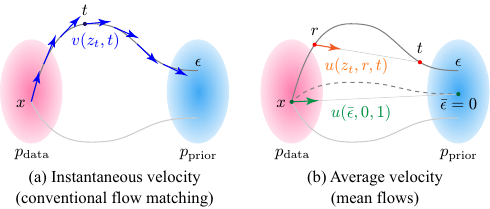}}
  \vspace{-2mm}
  \caption{Comparison of (a) instantaneous velocity used in conventional flow matching and (b) average velocity used in mean flows.
    (a) \textit{Instantaneous velocity} $v(z_t, t)$ (blue arrow) represents the tangent direction of the path for a single time step $t$.
    (b) \textit{Average velocity} $u(z_t, r, t)$ (orange arrow) aligns with the displacement between two time steps $r$ and $t$.
    In \textit{MeanVoiceFlow}, a \textit{zero-input constraint} is imposed on $u(\bar{\epsilon}, 0, 1)$ (green arrow), the average velocity for a zero-input sample $\bar{\epsilon} = 0$, using a \textit{structural margin reconstruction loss} to moderately guide learning.}
  \vspace{-3mm}
  \label{fig:teaser}
\end{figure}

To address these limitations, we propose \textit{MeanVoiceFlow}, a novel \textit{one-step} nonparallel VC model based on mean flows~\cite{ZGengNeurIPS2025}, which can be trained \textit{from scratch} without requiring pretraining or distillation.
This is enabled by replacing \textit{instantaneous velocity}, which is used in conventional flow matching~\cite{YLipmanICLR2023,XLiuICLR2023,MAlbergoICLR2023}, with \textit{average velocity}.
As illustrated in Fig.~\ref{fig:teaser}, conventional flow matching uses the instantaneous velocity $v(z_t, t)$ (Fig.~\ref{fig:teaser}(a)), which represents the tangent direction of the path at $t$.
The path integral is approximated by discretizing the path and integrating these velocities using an ordinary differential equation (ODE) solver (e.g., Euler method).
However, such discretization can introduce integration errors, particularly when using coarse steps.
In contrast, mean flows use the average velocity $u(z_t, r, t)$ (Fig.~\ref{fig:teaser}(b)), which corresponds to the displacement between two time steps $r$ and $t$.
This allows the path integral to be computed directly using this velocity, without numerical approximation, thereby enabling more accurate one-step inference.

However, a key challenge in training the average velocity is that computing the target velocity requires its derivative, which can be inaccurate during training.
A simple solution involves regularizing the input--output behavior of the model using an element-wise reconstruction loss (e.g., mean squared error loss).
However, overly strong constraints often cause statistical averaging, leading to perceptually buzzy outputs.
Consequently, we introduce a \textit{structural margin reconstruction loss} as a \textit{zero-input constraint}, which moderates the regularization by (1) using a structural similarity index measure (SSIM)~\cite{ZWangTIP2004}-based loss instead of a direct element-wise loss, (2) applying a margin to ignore high-quality samples, and (3) restricting the constraint to zero-input samples (green arrow in Fig.~\ref{fig:teaser}(b)).

Furthermore, unlike previous studies (e.g.,~\cite{TKanekoIS2024}) that feed a mixture of noise and source data only during inference, we apply it during training as well.
This strategy, termed \textit{conditional diffused-input training}, enables effective use of source information while maintaining consistency between training and inference.

\textit{MeanVoiceFlow} was experimentally evaluated for nonparallel any-to-any (i.e., \textit{zero-shot}) VC tasks.
The results verified the efficacy of the proposed techniques and demonstrated that \textit{MeanVoiceFlow} can achieve performance comparable to that of previous multi-step and distillation-based models, even when trained \textit{from scratch}, without requiring pretraining or distillation.

The remainder of this paper is organized as follows:
Section~\ref{sec:preliminaries} reviews flow matching and mean flows, which forms the basis of \textit{MeanVoiceFlow}.
Section~\ref{sec:meanvoiceflow} introduces the proposed \textit{MeanVoiceFlow}.
Section~\ref{sec:experiments} presents experimental results.
Section~\ref{sec:conclusion} concludes the paper and discusses directions for future research.

\section{Preliminaries}
\label{sec:preliminaries}

\subsection{Flow matching}
\label{subsec:flow_matching}

Flow matching~\cite{YLipmanICLR2023,XLiuICLR2023,MAlbergoICLR2023} is a technique for learning to match the flows between two probabilistic distributions.
Given data $x \sim p_{\mathrm{data}}(x)$ (specifically, a log-mel spectrogram in our task) and prior $\epsilon \sim p_{\mathrm{prior}}(\epsilon)$ (typically, $p_{\mathrm{prior}} = \mathcal{N}(0, 1)$), a flow path is defined as follows:
\begin{flalign}
  \label{eq:cfm_path}
  z_t = (1 - t) x + t \epsilon
\end{flalign}
for time step $t \in [0, 1]$.
The corresponding conditional velocity is defined as $v_t = v_t(z_t | x) = \frac{dz_t}{dt} = \epsilon - x$.
To ensure that $z_t$ does not correspond to multiple $(x, \epsilon)$ pairs, the marginal instantaneous velocity (Fig.~\ref{fig:teaser}(a)) is introduced as $v(z_t, t) = \mathbb{E}_{p_t(v_t | z_t)} [v_t]$.

\smallskip\noindent\textbf{Training.}
A neural network $v_{\theta}(z_t, t)$, parameterized by $\theta$, is trained to approximate $v(z_t, t)$.
However, a flow matching loss defined as $\mathcal{L}_{\mathrm{FM}} = \mathbb{E} [d(v_{\theta}(z_t, t), v(z_t, t))]$, where $d(\cdot, \cdot)$ denotes a distance metric, is intractable because computing $v(z_t, t)$ requires marginalization over $(x, \epsilon)$.
Alternatively, $v_t(z_t | x)$ is used as the target, and a conditional flow matching loss is minimized:
\begin{flalign}
  \label{eq:cfm_loss}
  \mathcal{L}_{\mathrm{CFM}} = \mathbb{E} [d(v_{\theta}(z_t, t), v_t(z_t | x))].
\end{flalign}
It has been shown in~\cite{YLipmanICLR2023} that minimizing $\mathcal{L}_{\mathrm{CFM}}$ is theoretically equivalent to minimizing the original flow matching loss $\mathcal{L}_{\mathrm{FM}}$.

\smallskip\noindent\textbf{Sampling.}
A sample is generated from $z_1 = \epsilon \sim p_{\mathrm{prior}}(\epsilon)$ by solving the ODE $\frac{dz_t}{dt} = v_{\theta}(z_t, t)$.
The solution can be expressed as:
\begin{flalign}
  \label{eq:fm_path}
  z_r = z_t - \int_r^t v_{\mathrm{\theta}}(z_r, \tau)d\tau
\end{flalign}
for another time step $r < t$.
However, the exact computation of this integral is intractable in practice.
Therefore, flow matching typically approximates it by discretizing time and solving the resulting ODE using numerical methods such as Euler integration.

\subsection{Mean flows}
\label{subsec:mean_flows}

Discretization using the instantaneous velocity $v(z_t, t)$ (Fig.~\ref{fig:teaser}(a)), as discussed in the previous section, can result in significant errors, particularly with few time steps.
Consequently, mean flows~\cite{ZGengNeurIPS2025} introduce the average velocity (Fig.~\ref{fig:teaser}(b)), defined as $u(z_t, r, t) = \frac{1}{t - r} \int_r^t v(z_{\tau}, \tau) d\tau$, which corresponds to the displacement between two times steps $r$ and $t$, inherently incorporating integration over time.
By multiplying both sides by $t - r$, differentiating with respect to $t$, and rearranging terms, we obtain a mean flow identity:
\begin{flalign}
  \label{eq:mf_identity}
  u(z_t, r, t) = v(z_t, t) - (t - r) \frac{d}{dt} u(z_t, r, t).
\end{flalign}
The total derivative $\frac{d}{dt} u(z_t, r, t)$ can be expanded in partial derivatives as $\frac{d}{dt} u(z_t, r, t) = v(z_t, t) \partial_z u + \partial_t u$.
This expansion can be further rewritten as the Jacobian-vector product (JVP) between the Jacobian matrix of $u$, that is, $[\partial_z u, \partial_r u, \partial_t u]$, and the tangent vector $[v, 0, 1]$.
This JVP can be efficiently computed using the $\texttt{jvp}$ interface available in modern deep learning libraries.

\smallskip\noindent\textbf{Training.}
A neural network $u_{\theta}(z_t, r, t)$ is optimized to satisfy the mean flow identity (Eq.~\ref{eq:mf_identity}) by minimizing a mean flow loss:
\begin{flalign}
  \label{eq:mf_loss}
  \mathcal{L}_{\mathrm{MF}} = \mathbb{E} [d(u_{\theta}(z_t, r, t), \mathrm{sg}(u_{\mathrm{tgt}}))],
\end{flalign}
where the target velocity $u_{\mathrm{tgt}}$ is defined as:
\begin{flalign}
  \label{eq:mf_target}
  u_{\mathrm{tgt}} = v_t - (t - r) (v_t \partial_z u_{\theta} + \partial_t u_{\theta}),
\end{flalign}
and $\mathrm{sg}$ denotes a stop-gradient operation.
Following the mean flow study~\cite{ZGengNeurIPS2025}, in the experiments, we adopt the adaptively weighted loss~\cite{ZGengICLR2025} as the distance metric $d$ in Eq.~\ref{eq:mf_loss}, defined as $d(a, b) = \frac{\| a - b \|_2^2}{\mathrm{sg}(\| a - b \|_2^2 + 10^{-3})}$.
Additionally, we randomly set $r = t$ with a probability of $0.75$ to stabilize training, thereby blending the training of the instantaneous and average velocity fields.

\smallskip\noindent\textbf{Sampling.}
A sample is generated from $z_1 = \epsilon \sim p_{\mathrm{prior}}$ using
\begin{flalign}
  \label{eq:mf_path}
  z_r = z_t - (t - r) u_{\theta}(z_t, r, t),
\end{flalign}
where the time integral, which is used in flow matching (Eq.~\ref{eq:fm_path}), is replaced with the product of the time duration and average velocity, $(t - r) u_{\theta}(z_t, r, t)$, which reduces discretization errors.
In the one-step sampling case, Eq.~\ref{eq:mf_path} simplifies to
\begin{flalign}
  \label{eq:mf_path_1step}
  z_0 = z_1 - u_{\theta}(z_1, 0, 1).
\end{flalign}

\section{MeanVoiceFlow}
\label{sec:meanvoiceflow}

\subsection{Extension to conditional generation}
\label{subsec:conditional_generation}

Thus far, we have described the formulation for unconditional generation.
However, VC aims to modify speaker identity while preserving linguistic content.
Hence, we adopt conditional generation that incorporates speaker and linguistic information.
Formally, we extend average velocity $u(z_t, r, t)$ introduced in Section~\ref{subsec:mean_flows} to its conditional form $u(z_t, r, t, s, c)$, where $s$ and $c$ denote the speaker embedding (e.g., extracted via a speaker encoder~\cite{YJiaNeurIPS2018}) and content embedding (e.g., extracted via a bottleneck feature extractor~\cite{SLiuTASPL2021}), respectively.
This extension is fully compatible with the mean flow algorithm described above and only requires replacing the average velocity with its conditional form.
For clarity, we omit $s$ and $c$ if this omission does not affect understanding.

\subsection{Zero-input constraint}
\label{subsec:zero_input}

As defined in Eq.~\ref{eq:mf_target}, the target velocity $u_{\mathrm{tgt}}$ in mean flows depends on the derivative of $u_{\theta}$, which may be inaccurate when $u_{\theta}$ is under training, leading to instability.
A simple solution is to regularize the input--output behavior of the model using a direct element-wise reconstruction loss $\mathcal{L}_{\mathrm{rec}} = \mathbb{E} \| \hat{x} - x \|_p^p$, where $\hat{x}$ denotes the output $z_0$ from Eq.~\ref{eq:mf_path_1step}, $x$ denotes the ground-truth data, and $p$ typically takes a value of $1$ or $2$.
However, such a strong constraint can cause over-smoothing (i.e., statistical averaging), as often observed in variational autoencoder-based models~\cite{DKingmaICLR2014}.
Consequently, we introduce a \textit{structural margin reconstruction loss} as a \textit{zero-input constraint}:
\begin{flalign}
  \mathcal{L}_{\mathrm{zerorec}} = \mathbb{E} [\max(1 - \mathrm{SSIM}(\bar{x}, x), m)],
\end{flalign}
where $\mathrm{SSIM}$ denotes the SSIM similarity~\cite{ZWangTIP2004}, $\bar{x}$ denotes the output $z_0$ from Eq.~\ref{eq:mf_path_1step} with zero input $z_1 = \bar{\epsilon} = 0$, and $m$ is a margin, empirically set to $0.3$.
To avoid over-regularization in $\mathcal{L}_{\mathrm{rec}}$, we adopt the following three strategies for $\mathcal{L}_{\mathrm{zerorec}}$:
(1) \textbf{Structural comparison:} We use an SSIM-based loss instead of a direct element-wise loss, focusing on structural rather than point-wise similarities.
(2) \textbf{Margin-based relaxation:} A margin $m$ is introduced to prevent penalizing high-quality samples, which are trained only with $\mathcal{L}_{\mathrm{MF}}$.
(3) \textbf{Selective application:} The constraint is applied only to zero-input samples (i.e., the center of $p_{\mathrm{prior}}$, marked by the green arrow in Fig.~\ref{fig:teaser}(b)).
Accordingly, the final objective is defined as:
\begin{flalign}
  \mathcal{L}_{\mathrm{MVF}} = \mathcal{L}_{\mathrm{MF}} + \lambda \mathcal{L}_{\mathrm{zerorec}},
\end{flalign}
where $\lambda$ is a weighting hyperparameter empirically set to $1$.

\subsection{Conditional diffused-input training}
\label{subsec:conditional_training}

\begin{figure}[t]
  \centering
  \centerline{\includegraphics[width=\columnwidth]{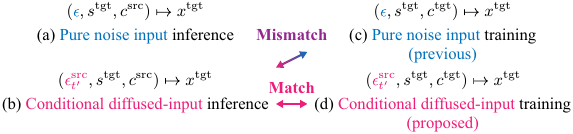}}
  \vspace{-4mm}
  \caption{Comparison of input types during inference and training.
    Previous studies (e.g.,~\cite{TKanekoIS2024}) use input type (c) during training and (b) during inference, causing a training--inference mismatch.
    In contrast, the proposed method uses (d) during training and (b) during inference, effectively eliminating this mismatch.}
  \vspace{-4mm}
  \label{fig:input_comparison}
\end{figure}

As VC aims to convert voice while preserving linguistic content, it is essential to utilize the source input effectively, without omission or redundancy.
Consequently, prior diffusion-based VC methods (e.g.,~\cite{TKanekoIS2024}) have proposed feeding diffused source data $\epsilon_{t'}^{\mathrm{src}} = (1 - t') x^{\mathrm{src}} + t' \epsilon$\footnote{More specifically, in a diffusion-based VC model~\cite{TKanekoIS2024}, the diffused source data is constructed as $\epsilon_{S_K}^{\mathrm{src}} = \sqrt{\bar{\alpha}_{S_K}} x^{\mathrm{src}} + \sqrt{1 - \bar{\alpha}_{S_K}} \epsilon$ to preserve variance (see Eq.~9 in~\cite{TKanekoIS2024} for details).
  In this study, we modify this formulation to match the linear mixture defined in Eq.~\ref{eq:cfm_path}.} into the model during inference (Fig.~\ref{fig:input_comparison}(b)), rather than using pure noise $\epsilon \sim \mathcal{N}(0, 1)$ (Fig.~\ref{fig:input_comparison}(a)), where $x^{\mathrm{src}}$ denotes the source data and $t' \in [0, 1]$ is a mixing ratio, empirically set to $0.95$ during inference.
We use superscripts $\mathrm{src}$ and $\mathrm{tgt}$ to denote data related to the source and target speakers, respectively.
However, this method introduces a training--inference mismatch: the model is trained only on pure noise $\epsilon$ (Fig.~\ref{fig:input_comparison}(c)), but receives the diffused source data $\epsilon_{t'}^{\mathrm{src}}$ during inference (Fig.~\ref{fig:input_comparison}(b)).
This discrepancy can degrade performance, since the model is not exposed during training to the input distribution encountered at inference time.

Consequently, we adopt \textit{conditional diffused-input training}, feeding diffused source data $\epsilon_{t'}^{\mathrm{src}}$ during training as well (Fig.~\ref{fig:input_comparison}(d)).
Specifically, $\epsilon_{t'}^{\mathrm{src}}$ is constructed such that the speaker identity differs from the target, simulating the actual conversion process.
However, in nonparallel VC, paired source--target data are unavailable, making it challenging to obtain such diffused source data directly.
To overcome this, we synthesize $\epsilon_{t'}^{\mathrm{src}}$ using the model itself.
Formally, given $x^{\mathrm{tgt}}$, we approximately define the diffused source data based on Eq.~\ref{eq:mf_path} as $\hat{\epsilon}_{t'}^{\mathrm{src}} = \mathrm{sg}(z_1 - (1 - t') u_{\theta}(z_1, t', 1, s^{\mathrm{src}}, c^{\mathrm{tgt}}))$, where $z_1 = \epsilon \sim \mathcal{N}(0, 1)$, $s^{\mathrm{src}}$ is obtained by shuffling $s^{\mathrm{tgt}}$ within the batch, and the stop-gradient operation $\mathrm{sg}$ is applied to simplify learning.
We then replace $\epsilon$ and $x$ in Section~\ref{sec:preliminaries} with $\hat{\epsilon}_{t'}^{\mathrm{src}}$ and $x^{\mathrm{tgt}}$, respectively.
In practice, we provide $t'$ as an additional conditional input to help the model distinguish input types, with negligible computational overhead.
Furthermore, to stabilize training, we use $\hat{\epsilon}_{t'}^{\mathrm{src}}$ for half of the batch and pure noise $\epsilon \sim \mathcal{N}(0, 1)$ for the other half.

\section{Experiments}
\label{sec:experiments}

\subsection{Experimental setup}
\label{subsec:experimental_setup}

\textbf{Data.}
We conducted experiments on nonparallel any-to-any (i.e., \textit{zero-shot}) VC tasks to evaluate \textit{MeanVoiceFlow}.
The experimental design followed that of a previous study on fast diffusion-based VC (\textit{FastVoiceGrad}~\cite{TKanekoIS2024}) which was used as the baseline model.
The main evaluations (Section~\ref{subsec:component_analysis} and \ref{subsec:comparative_experiments}) were performed on the VCTK dataset~\cite{JYamagishiVCTK2019}, which contains the speech recordings of 110 English speakers.
To examine dataset dependency (Section~\ref{subsec:versatility_analysis}), we additionally used the train-clean subset of the LibriTTS dataset~\cite{HZenIS2019}, which contains the speech recordings of 1,151 English speakers.
We excluded ten evaluation speakers and ten evaluation sentences from the training set to simulate unseen-to-unseen VC scenarios.
All audio clips were downsampled to 22.05 kHz, and 80-dimensional log-mel spectrograms were extracted using an FFT size of 1024, hop size of 256, and window size of 1024.
These spectrograms were used as the conversion targets, i.e., as $x$.

\smallskip\noindent\textbf{Implementation.}
To isolate the effect of training methods, we adopted the same network architectures as the baseline~\cite{TKanekoIS2024}.
The average velocity $u_{\theta}$, corresponding to the noise predictor $\epsilon_{\theta}$ in~\cite{TKanekoIS2024}, was implemented using a U-Net~\cite{ORonnebergerMICCAI2015} with 12 convolution layers, 512 hidden channels, two downsampling and upsampling stages, gated linear units~\cite{YDauphinICML2017}, and weight normalization~\cite{TSalimansNIPS2016}.
Speaker and content embeddings $s$ and $c$ were extracted using a speaker encoder~\cite{YJiaNeurIPS2018} and a bottleneck feature extractor~\cite{SLiuASRU2021}, respectively.
Waveforms were synthesized from log-mel spectrograms using HiFi-GAN V1~\cite{JKongNeurIPS2020}.
Models were trained using the Adam optimizer~\cite{DPKingmaICLR2015} with a batch size of $32$, learning rate of $0.0002$, $\beta_1$ of $0.5$, and $\beta_2$ of $0.9$.
Training was performed for $500$ epochs using a cosine learning rate schedule with a linear warm-up over the first 10k steps.
During training, $t$, $r$, and $t'$ were sampled from a logit-normal distribution~\cite{PEsserICML2024} by applying the sigmoid function to samples from $\mathcal{N}(0, 1)$; 
the larger of $t$ and $r$ was assigned to $t$, and the smaller to $r$.

\smallskip\noindent\textbf{Evaluation metrics.}
For efficient model comparison, we primarily relied on objective metrics and performed subjective evaluations for the critical comparisons (Section~\ref{subsec:comparative_experiments}).
The five objective metrics were as follows:
(1) \texttt{pMOS$_{\mathrm{s}}$}$\uparrow$: Predicted mean opinion score (MOS) for \textit{synthesized} speech using UTMOS~\cite{TSaekiIS2022}.
(2) \texttt{pMOS$_{\mathrm{n}}$}$\uparrow$: Predicted MOS for \textit{noise-suppressed} speech using DNSMOS~\cite{CReddyICASSP2021}.
(3) \texttt{pMOS$_{\mathrm{v}}$}$\uparrow$: Predicted MOS for \textit{voice-converted} speech using DNSMOS Pro~\cite{FCumlinIS2024} on VCC2018~\cite{JLorenzoOdyssey2018}.
(4) \texttt{CER}$\downarrow$: Character error rate using Whisper-large-v3~\cite{ARadfordICML2023}, measuring \textit{speech intelligibility}.
(5) \texttt{SECS}$\uparrow$: Speaker embedding cosine similarity using WavLM Base+~\cite{SChenJSTSP2022}, measuring \textit{speaker similarity}.
We computed these metrics for 8,100 unique speaker--sentence pairs from the evaluation set.
Audio samples are available online.\footnoteref{foot:samples}

\subsection{Component analysis}
\label{subsec:component_analysis}

\textbf{Analysis of zero-input constraint (Section~\ref{subsec:zero_input}).}
We evaluated the effectiveness of three techniques for zero-input constraint: (1) \textit{structural comparison}, (2) \textit{margin-based relaxation}, and (3) \textit{selective application}.
To isolate the effect of the loss function, conditional diffused-input training was not applied in this experiment.
As shown in Table~\ref{tab:analysis_zero_input}, the baseline (\texttt{A}) used only $\mathcal{L}_{\mathrm{MF}}$ (Eq.~\ref{eq:mf_loss}).
Configurations (\texttt{B})--(\texttt{D}) added reconstruction losses to this baseline.
Element-wise losses in (\texttt{B}) and (\texttt{C}) improved \texttt{pMOS$_{\mathrm{s}}$} but degraded \texttt{pMOS$_{\mathrm{n}}$} and \texttt{pMOS$_{\mathrm{v}}$} owing to over-smoothing.
Using an SSIM loss in (\texttt{D}) mitigated this issue, though \texttt{pMOS$_{\mathrm{s}}$} remained low.
Introducing a margin in (\texttt{E}) further improved all scores by softening the constraint.
However, applying the constraint to all inputs in (\texttt{F}) caused over-smoothing again and reduced scores, particularly in \texttt{pMOS$_{\mathrm{v}}$}.
These results highlight the importance of incorporating the reconstruction loss and validate the proposed configuration (\texttt{E}).

\begin{table}[h]
  \vspace{-5mm}
  \caption{Analysis of zero-input constraint.
    Performance changes were analyzed while varying \texttt{metric}, \texttt{margin}, and \texttt{input} type in the constraint.
    (\texttt{E}) denotes the proposed configuration.}
  \vspace{0.5mm}
  \label{tab:analysis_zero_input}
  \newcommand{\spm}[1]{{\tiny$\pm$#1}}
  \setlength{\tabcolsep}{2pt}
  \centering
  {\fontsize{6.5pt}{7pt}\selectfont
    \begin{tabularx}{\columnwidth}{ccccCCCCC}
      \toprule
      & \texttt{Metric} & \texttt{Margin} & \texttt{Input}
      & \texttt{pMOS$_{\mathrm{s}}$}$\uparrow$ & \texttt{pMOS$_{\mathrm{n}}$}$\uparrow$ & \texttt{pMOS$_{\mathrm{v}}$}$\uparrow$ & \texttt{CER}$\downarrow$ & \texttt{SECS}$\uparrow$
      \\ \midrule
      (\texttt{A}) & -- & -- & --
      & 3.72 & \third{3.73} & 4.00 & \second{1.3} & \second{0.882}
      \\
      (\texttt{B}) & L1 & -- & Zero
      & \third{3.81} & 3.65 & 3.96 & \first{1.2} & \second{0.882}
      \\
      (\texttt{C}) & L2 & -- & Zero
      & 3.80 & 3.66 & 3.98 & \first{1.2} & \second{0.882}
      \\
      (\texttt{D}) & SSIM & -- & Zero
      & 3.79 & \first{3.77} & \second{4.05} & \second{1.3} & \second{0.882}
      \\ \midrule
      \textstrong{(\texttt{E})} & \textstrong{SSIM} & \textstrong{$\checkmark$} & \textstrong{Zero}
      & \first{3.90} & \first{3.77} & \first{4.08} & \first{1.2} & \first{0.883}
      \\ \midrule
      (\texttt{F}) & SSIM & $\checkmark$ & All
      & \second{3.89} & \second{3.76} & \third{4.03} & \second{1.3} & \first{0.883}
      \\ \bottomrule
    \end{tabularx}
  }
  \vspace{-2.5mm}
\end{table}

\begin{figure}[h]
  \centering
  \vspace{-2mm}
  \centerline{\includegraphics[width=0.98\columnwidth]{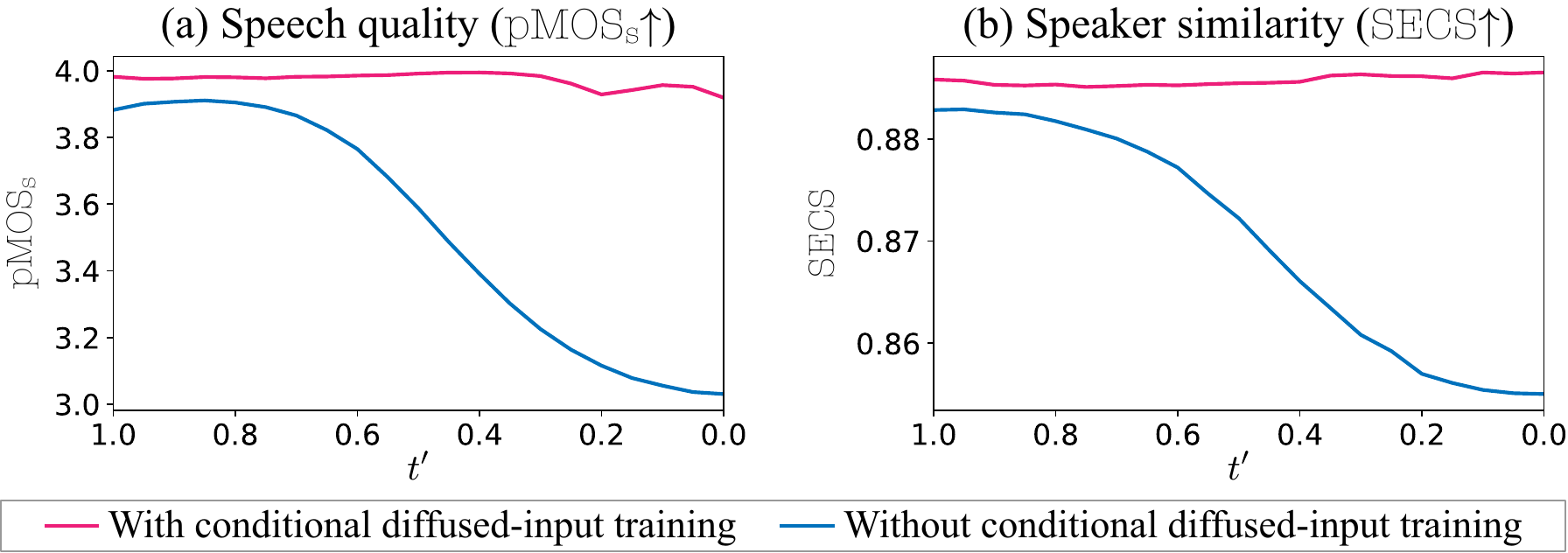}}
  \vspace{-3mm}
  \caption{Analysis of conditional diffused-input training.
    Conditional diffused-input training (pink line) enhances both robustness to the mixing ratio $t'$ and peak performance.}
  \vspace{-6mm}
  \label{fig:analysis_conditional_training}
\end{figure}

\smallskip\noindent\textbf{Analysis of conditional diffused-input training (Section~\ref{subsec:conditional_training}).}
We evaluated the effectiveness of conditional diffused-input training by comparing models trained with and without it while varying the mixing ratio $t'$ during inference.
As shown in Fig.~\ref{fig:analysis_conditional_training}, the proposed training improved robustness to $t'$ and peak performance for both \texttt{pMOS$_{\mathrm{s}}$} and \texttt{SECS}.
Similar trends were observed for \texttt{pMOS$_{\mathrm{n}}$} and \texttt{pMOS$_{\mathrm{v}}$} while \texttt{CER} remained nearly constant.
These results validate the utility of conditional diffused-input training.

\subsection{Comparisons with prior models}
\label{subsec:comparative_experiments}

We compared \textit{MeanVoiceFlow} with previous models to assess its relative efficacy.
We maintained consistent network architectures and varied only the training procedures to isolate the impact of training strategies.
We evaluated six baselines:
\textit{VoiceGrad-DM-1/30}~\cite{HKameokaTASLP2024}: Diffusion~\cite{JHoNeurIPS2020}-based VoiceGrad, evaluated with 1 and 30 NFEs (number of function evaluations, i.e., reverse diffusion steps).
The smaller the NFE, the faster the inference.
\textit{VoiceGrad-FM-1/30}~\cite{HKameokaTASLP2025}: Flow matching~\cite{XLiuICLR2023}-based VoiceGrad, evaluated with 1 and 30 NFEs.
\textit{FastVoiceGrad}~\cite{TKanekoIS2024}: One-step diffusion-based VC model with distillation and adversarial training~\cite{ASauerECCV2024}.
\textit{FastVoiceGrad+}~\cite{TKanekoIS2025b}: FastVoiceGrad with an enhanced discriminator.
We also evaluated \textit{DiffVC-30}~\cite{VPopovICLR2022}, a commonly used baseline, and \textit{ground-truth} speech as reference anchors.
For comprehensive evaluation, we performed MOS tests on $90$ speaker--sentence pairs per model.
Naturalness was assessed using a five-point scale (\texttt{nMOS}: 1 = bad, 2 = poor, 3 = fair, 4 = good, and 5 = excellent) and speaker similarity was assessed using a four-point scale (\texttt{sMOS}: 1 = different (sure), 2 = different (not sure), 3 = same (not sure), and 4 = same (sure)).
The tests were performed online with ten participants, and over $200$ responses were collected for each model.
As summarized in Table~\ref{tab:comparative_experiments}, \textit{MeanVoiceFlow} outperforms one-step models trained from scratch (\textit{VoiceGrad-DM-1} and \textit{VoiceGrad-FM-1}), and achieves performance comparable to that of multi-step models (\textit{VoiceGrad-DM-30} and \textit{VoiceGrad-FM-30}) and one-step models enhanced by distillation and adversarial training (\textit{FastVoiceGrad} and \textit{FastVoiceGrad+}).
Notably, \textit{MeanVoiceFlow} achieves this strong performance \textit{without requiring pretraining or distillation}.

\begin{table}[h]
  \vspace{-4mm}
  \caption{Comparisons with prior models.
    $^\dag$ indicates models that require additional modules (a pretrained teacher and a pretrained feature extractor for the discriminator) and extra training procedures (distillation and adversarial training).
    For \texttt{nMOS} and \texttt{sMOS}, $^*$ indicates a statistically significant difference from \textit{MeanVoiceFlow} based on the Mann--Whitney U test ($p < 0.05$).
    Values following $\pm$ represent 95\% confidence intervals.}
  \vspace{0.5mm}
  \label{tab:comparative_experiments}
  \newcommand{\spm}[1]{{\tiny$\pm$#1}}
  \setlength{\tabcolsep}{0.7pt}
  \centering
  {\fontsize{6.5pt}{7pt}\selectfont
    \begin{tabularx}{\columnwidth}{lcllcccccc}
      \toprule
      \multicolumn{1}{c}{\texttt{Model}} & \texttt{NFE}$\downarrow$
      & \multicolumn{1}{c}{\texttt{nMOS}$\uparrow$} & \multicolumn{1}{c}{\texttt{sMOS}$\uparrow$} & \texttt{pMOS$_{\mathrm{s}}$}$\uparrow$ & \texttt{pMOS$_{\mathrm{n}}$}$\uparrow$ & \texttt{pMOS$_{\mathrm{v}}$}$\uparrow$ & \texttt{CER}$\downarrow$ & \texttt{SECS}$\uparrow$
      \\ \midrule
      Ground truth & --
      & \: 4.26\spm{0.10}$^*$ & \: 3.62\spm{0.08}$^*$ & 4.14 & 3.75 & 4.05 & 0.1 & 0.940
      \\
      DiffVC & 30
      & \: 3.59\spm{0.11}$^*$ & \: 2.55\spm{0.13}$^*$ & 3.76 & 3.75 & 4.03 & 5.4 & 0.880
      \\ \midrule
      VoiceGrad-DM & 1
      & \: 2.77\spm{0.09}$^*$ & \: 2.48\spm{0.13}$^*$ & 3.72 & 3.68 & 3.99 & 1.4 & 0.884
      \\
      VoiceGrad-DM & 30
      & \thirdl{\: 3.79\spm{0.11}} & \: 2.70\spm{0.13}$^*$ & 3.88 & \third{3.77} & 4.02 & 1.4 & \third{0.885}
      \\
      VoiceGrad-FM & 1
      & \: 3.14\spm{0.11}$^*$ & \: 2.60\spm{0.13}$^*$ & 3.81 & 3.69 & 4.01 & \first{1.1} & \third{0.885}
      \\
      VoiceGrad-FM & 30
      & \thirdl{\: 3.79\spm{0.10}} & \thirdl{\: 2.92\spm{0.12}} & 3.88 & \first{3.79} & \second{4.05} & \first{1.1} & \third{0.885}
      \\ \midrule
      FastVoiceGrad$^\dag$ & 1
      & \: 3.73\spm{0.09}$^*$ & \secondl{\: 2.93\spm{0.11}} & \third{3.96} & \third{3.77} & \third{4.04} & \third{1.3} & \first{0.888}
      \\
      FastVoiceGrad+$^\dag$ & 1
      & \secondl{\: 3.81\spm{0.10}} & \firstl{\: 2.99\spm{0.13}} & \first{3.99} & \first{3.79} & 4.03 & \second{1.2} & \first{0.888}
      \\ \midrule
      \textstrong{\textit{MeanVoiceFlow}} & 1
      & \firstl{\: 3.87\spm{0.09}} & \thirdl{\: 2.92\spm{0.13}} & \second{3.98} & \second{3.78} & \first{4.10} & \second{1.2} & \second{0.886}
      \\ \bottomrule
    \end{tabularx}
  }
  \vspace{-4mm}
\end{table}

\subsection{Versatility analysis}
\label{subsec:versatility_analysis}

To examine dataset dependency, we evaluated \textit{VoiceGrad-DM-1/30}, \textit{VoiceGrad-FM-1/30}, and \textit{MeanVoiceFlow}, which were trained under similar conditions (i.e., \textit{from scratch}), on the LibriTTS dataset~\cite{HZenIS2019}.
As summarized in Table~\ref{tab:versatility_analysis}, a similar trend was observed:
\textit{MeanVoiceFlow} outperforms one-step models (\textit{VoiceGrad-DM-1} and \textit{VoiceGrad-FM-1}) and achieves performance comparable to that of multi-step models (\textit{VoiceGrad-DM-30} and \textit{VoiceGrad-FM-30}).
Audio samples are available online.\footnoteref{foot:samples}

\begin{table}[h]
  \vspace{-4mm}
  \caption{Versatility analysis on LibriTTS.
    Models trained under similar conditions (i.e., \textit{from scratch}) were compared.}
  \vspace{0.5mm}
  \label{tab:versatility_analysis}
  \newcommand{\spm}[1]{{\tiny$\pm$#1}}
  \setlength{\tabcolsep}{3pt}
  \centering
  {\fontsize{6.5pt}{7pt}\selectfont
    \begin{tabularx}{\columnwidth}{lCCCCCCC}
      \toprule
      \multicolumn{1}{c}{\texttt{Model}} & \texttt{NFE}$\downarrow$
      & \texttt{pMOS$_{\mathrm{s}}$}$\uparrow$ & \texttt{pMOS$_{\mathrm{n}}$}$\uparrow$ & \texttt{pMOS$_{\mathrm{v}}$}$\uparrow$ & \texttt{CER}$\downarrow$ & \texttt{SECS}$\uparrow$
      \\ \midrule
      VoiceGrad-DM & 1
      & 3.20 & 3.32 & 3.26 & \first{1.1} & \third{0.873}
      \\
      VoiceGrad-DM & 30
      & \second{3.81} & \second{3.75} & \second{3.58} & \second{1.2} & 0.865
      \\
      VoiceGrad-FM & 1
      & 3.22 & 3.38 & 3.28 & \first{1.1} & \second{0.875}
      \\
      VoiceGrad-FM & 30
      & \third{3.77} & \first{3.77} & \third{3.38} & \third{1.3} & 0.866
      \\ \midrule
      \textstrong{\textit{MeanVoiceFlow}} & 1
      & \first{3.93} & \third{3.70} & \first{3.70} & \first{1.1} & \first{0.879}
      \\ \bottomrule
    \end{tabularx}
  }
  \vspace{-4mm}
\end{table}

\section{Conclusion}
\label{sec:conclusion}

We proposed \textit{MeanVoiceFlow}, a novel \textit{one-step} nonparallel VC model built upon mean flows, which can be trained \textit{from scratch} without requiring pretraining or distillation.
To enhance its performance, we introduced \textit{zero-input constraint} and \textit{conditional diffused-input training}.
Experiments demonstrate the effectiveness of these techniques and show that the proposed method achieves performance comparable to that of previous multi-step and distillation-based models.
The application of mean flows in speech remains underexplored; thus, extending the proposed approach to other speech-related tasks is a promising direction for future work.

\clearpage
\bibliographystyle{IEEEbib}
{\scriptsize\bibliography{refs}}

\end{document}